\documentclass[fleqn,usenatbib]{mnras}

\usepackage{newtxtext,newtxmath}
\usepackage{tabularx}
\usepackage{enumitem}
\usepackage{multirow}
\usepackage{xcolor}
\usepackage[normalem]{ulem}
\usepackage{orcidlink}

\usepackage[T1]{fontenc}

\DeclareRobustCommand{\VAN}[3]{#2}
\let\VANthebibliography\thebibliography
\def\thebibliography{\DeclareRobustCommand{\VAN}[3]{##3}\VANthebibliography}
\usepackage{graphicx}	
\usepackage{amsmath}	
\usepackage[normalem]{ulem}

\title[A neutrino flare associated with ATLAS17jrp]{A neutrino flare candidate potentially associated with X-ray emission from tidal disruption event ATLAS17jrp}

\author[Rong-Lan Li et al.]{
Rong-Lan Li$^{1,2}$\orcidlink{0009-0002-9585-0210},
Chengchao Yuan$^{3}$\orcidlink{0000-0003-0327-6136},
Hao-Ning He$^{1,2}$\thanks{Corresponding author: Hao-Ning He, email: hnhe@pmo.ac.cn}\orcidlink{0000-0002-8941-9603},
Yun Wang$^{1}$\orcidlink{0000-0002-8385-7848},
Ben-Yang Zhu$^{1,2}$\orcidlink{0000-0003-1511-5567},
Yun-Feng Liang$^{4}$\orcidlink{0000-0002-6316-1616},
\newauthor
Ning Jiang$^{2,5}$\orcidlink{0000-0002-7152-3621},
Da-Ming Wei$^{1,2}$\orcidlink{0000-0002-9758-5476}
\\
$^{1}$Key Laboratory of Dark Matter and Space Astronomy, Purple Mountain Observatory, Chinese Academy of Sciences, \\
Nanjing 210023, People's Republic of China\\
$^{2}$School of Astronomy and Space Sciences, University of Science and Technology of China, Hefei, 230026, People's Republic of China\\
$^{3}$Deutsches Elektronen-Synchrotron DESY, Platanenallee 6, 15738 Zeuthen, Germany\\
$^{4}$Laboratory for Relativistic Astrophysics, Department of Physics, Guangxi University, Nanning 530004, People's Republic of China\\
$^{5}$CAS Key laboratory for Research in Galaxies and Cosmology, Department of Astronomy, University of Science and Technology of China,\\
Hefei, 230026, People's Republic of China
}

\date{Accepted XXX. Received YYY; in original form ZZZ}

\pubyear{\the\year{}}

\begin{document}
\label{firstpage}
\pagerange{\pageref{firstpage}--\pageref{lastpage}}
\maketitle

\begin{abstract}
Tidal disruption events (TDEs), in which stars are disrupted by supermassive black holes, have been proposed as potential sources of high-energy neutrinos through hadronic interactions. X-ray-bright TDEs provide dense photon fields conducive to neutrino production via proton-photon ($p\gamma$) processes. We conducted a time-dependent unbinned likelihood analysis of ten years ($2008-2018$) of IceCube muon-track data, focusing on ten TDEs with confirmed X-ray detections during this period. We report a neutrino flare candidate spatially and temporally coincident with the TDE ATLAS17jrp, occurring 19 days after the onset of its X-ray activity and lasting for 56 days, with a post-trial $p$-value of 0.01. This significance is modest, representing a hint of an association. We illustrate the neutrino emission using a simple lepto-hadronic model, where X-ray photons serve as target fields. While this model can account for the neutrino data around 100 TeV, the low-energy neutrinos may imply contributions from an additional component. Although constrained by the sample size of X-ray-detected TDEs, these results underscore the need for high-cadence X-ray monitoring and future neutrino observatories to further explore the connection between TDEs and high-energy neutrinos.
\end{abstract}




\section{Introduction}
    In 2013, the IceCube Neutrino Observatory reported the first detection of astrophysical neutrinos in the TeV-PeV energy range, opening a new window into both astrophysics and particle physics \citep{IceCube:2013low, IceCube:2014stg}. Neutrinos interact only weakly with matter and magnetic fields during their propagation, offering unique insights into the extreme environments of their cosmic sources \citep{IceCube:2023agq}. This makes high-energy neutrinos key messengers in multimessenger astrophysics, particularly in identifying the sources of cosmic rays \citep{Ahlers:2017wkk, Guepin:2022qpl, Murase:2019tjj}. Such neutrinos are thought to originate from hadronic interactions of relativistic protons with ambient matter or photon fields, which also produce secondary $\gamma$-rays. Thus, identifying associations between neutrinos and electromagnetic counterparts is a promising approach to tracing the origins of astrophysical neutrinos.
    
    In recent years, the tidal disruption event (TDE) AT2019dsg and the TDE candidates AT2019fdr and AT2019aalc have been proposed associated with IceCube track alerts IC191001A \citep{AT2019dsg_ZTF}, IC200530A \citep{2019fdr_alert}, and IC191119A \citep{AT2019aalc_nu}, respectively, as their positions fall within the corresponding neutrino localization regions. \cite{Jiang:2023kbb} reported two possible associations between neutrino alerts and TDE candidates (SDSSJ1048+1228 and SDSSJ1649+2625) selected from the sample of mid-infrared outbursts in nearby galaxies (MIRONG) \citep{MIRONG_sample}. \cite{Yuan_2024a} suggested that the extremely luminous TDE candidate AT2021lwx may have a delayed neutrino counterpart. These findings have supported the hypothesis that TDEs could represent a promising class of high-energy neutrino sources \citep{Stein_2019,TDE_neutrino_2023}. 
    Recently, the IceCube Collaboration presented preliminary results for the updated catalog of alert tracks (IceCat-2), which features significantly improved directional reconstructions of neutrino events. Based on these improved localizations, the Collaboration reported that the three TDEs (AT2019dsg, AT2019fdr, and AT2019aalc) previously linked to neutrino alerts now lie well outside the revised 90$\%$ containment regions of their corresponding alerts, thus ruling out an association \citep{IceCat2}. This revision highlights the need for more systematic approaches to evaluating neutrino-TDE connections beyond isolated positional coincidences.
    
    TDEs are among the brightest transient phenomena observed in the optical, ultraviolet, and X-ray bands, with durations ranging from months to years. They occur when a star passes within the tidal radius of a supermassive black hole (SMBH) and is disrupted by tidal forces \citep{Rees:1988bf,Phinney_1989,Evans_1989ApJ,Cannizzo_1990ApJ}. Standard TDE models include key components such as accretion disks \citep{Hayasaki:2019kjy}, coronae \citep{2019fdr_alert}, sub-relativistic outflows or winds \citep{Fang_2020ApJ}, and potentially jets and dust tours \citep{Wang:2015mmh,Dai&Fang:2016gtz,Senno:2016bso,Lunardini:2016xwi,Dai:2018jbr,Zheng:2022afy,Yuan_2024c,Yuan_2024b}. High-energy astrophysical neutrinos are typically produced when accelerated cosmic rays interact with target photons or dense matter. In TDE environments, cosmic rays can be accelerated in coronae, jets, accretion disks, or winds \citep{Murase:2020lnu}, and may subsequently produce neutrinos via interactions with ambient radiation fields or surrounding matter. Possible target photons include thermal optical-ultraviolet (OUV), infrared (IR), and X-ray photons \citep{Winter_2023}. Due to the $\Delta$-resonance \citep{Waxman:1997ti}, relativistic protons interacting with photons of energies around keV typically produce neutrinos with energies near 10 TeV. This makes TDEs with X-ray observations plausible candidate sources of the TeV-PeV neutrinos observed by IceCube.
    
    Motivated by the uncertainties and limitations in previous candidate associations, we perform a time-dependent analysis to search for neutrino flares that are spatially and temporally correlated with the electromagnetic emission from X-ray-detected TDEs. In Section \ref{sec:data}, we introduce the IceCube muon-track dataset and the X-ray observations of TDEs collected between 2008 and 2018. In Section \ref{sec:analyse}, we describe the unbinned time-dependent analysis used to identify neutrino flares and present the results of the search. In Section \ref{model}, we outline a multi-messenger modeling framework that connects the observed neutrino signal with the broadband emission. We discuss the physical implications of our findings and summarize our conclusions in Section \ref{conclusion}.

\section{Data}\label{sec:data}
\subsection{The muon-track data from IceCube}\label{subsec:10yr data of IceCube}
    IceCube publicly released muon-track data spanning April 2008 to July 2018\footnote{\url{http://doi.org/DOI:10.21234/sxvs-mt83}}, comprising 1,134,450 events \citep{IceCube:2021xar}. This dataset is dominated by atmospheric muons and neutrinos produced in cosmic-ray air showers, which represent the main background for astrophysical neutrino searches. It has been extensively used in searches for point-like neutrino sources \citep{IceCube.time-integrated} and catalog analyses \citep{zhou2021,LiRonglan_2022}. Based on the instrument response, we categorized the data into five samples corresponding to different stages of IceCube construction, namely IC40, IC59, IC79, IC86-I, and IC86-II-VII. The numbers in each sample name denote the number of detector strings equipped with digital optical modules.
	
\subsection{TDEs with X-ray observations from 2008 to 2018}\label{subsec:TDE}
    To search for potential neutrino emission spatially coincident with TDEs that have X-ray observations, we compiled a sample of 10 TDEs observed between 2008 and 2018. This sample is drawn from recent X-ray TDE studies \citep{X-ray_TDE_Saxton_2021, Guolo:2023bds} and a single source analysis \citep{Wang&Jiang_2022_ATLAS17jrp}, which corresponds to the IceCube muon-track data period. Basic information on these sources is summarized in Table~\ref{table:TDE} and briefly described below:
    \begin{enumerate}[label=\arabic*., leftmargin=0pt, itemindent=!, align=left]
    \item ATLAS17jrp was initially detected by the Asteroid Terrestrial-impact Last Alert System \citep[ATLAS,][]{ATLAS_2018} and subsequently designated as AT2017gge\footnote{\url{https://www.wis-tns.org/object/2017gge}} on the Transient Name Server (TNS), where it was classified as a likely TDE \citep{2017TNS_ATLAS17jrp}. \cite{Wang&Jiang_2022_ATLAS17jrp} reported a detailed analysis of ATLAS17jrp across optical, X-ray, and infrared wavelengths, concluding that ATLAS17jrp is an extraordinary TDE. Its optical/UV light curves rise to the peak luminosity within approximately one month and then decay as $t^{-5/3}$, while the optical spectra show a blue continuum and very broad Balmer lines, consistent with other optical TDEs.
    
    \item ASASSN-14li was first discovered by the All-Sky Automated Survey for SuperNovae \citep[ASASSN,][]{ASASSN_Shappee_2014}. Subsequent analysis of ASASSN-14li revealed that its X-ray luminosity decreased more slowly than optical/UV emission, with X-rays becoming the dominant emission source around 40 days after the peak \citep{ASASSN-14li_2016}.
    
    \item ASASSN-15oi, another X-ray bright TDE discovered by ASASSN, exhibited atypical X-ray evolution. Initial observations indicated a low X-ray luminosity \citep{ASASSN-15oi}, but later monitoring revealed a dramatic brightening of nearly an order of magnitude before the X-ray emission faded again \citep{Gezari:2017qnq,ASASSN-15oi_2024}.
    
    \item OGLE16aaa was first identified through the OGLE survey \citep{OGLE16aaa_2017}. XMM-\textit{Newton} detected delayed X-ray emission $182\pm 5$ days after initial optical detection, with X-ray luminosity increasing by more than tenfold in one week \citep{OGLE16aaa_Shu_2020,OGLE16aaa_2020}. 
    
    \item PS18kh was detected by the Pan-STARRS Survey for Transients and designated as AT2018zr\footnote{\url{https://www.wis-tns.org/object/2018zr}} on the TNS. Observations showed that the source exhibited strong UV emission and weak soft X-rays \citep{AT2018zr_2019}.
    
    \item SDSS J120136.02+300305.5 (hereafter SDSS J1201+30) was identified as a TDE through an X-ray flare detected in the XMM-\textit{Newton} slew survey \citep{SDSSJ1201_Saxton_2012}. 
    
    \item 2MASX 07400785-8539307 (hereafter 2MASX 0740-85) is another TDE identified via XMM-\textit{Newton} X-ray observations \citep{MASX0740_Saxton_2017}.
    
    \item XMMSL2 J144605.0+685735 (hereafter XMMSL2 J1446+68) was detected by the XMM-\textit{Newton} slew survey \citep{XMMSL2J1446_Saxton_2019}. The X-ray flux was remained flat for approximately 100 days before decreasing by a factor of 100 over the following 500 days. 
    
    \item \textit{Swift} J164449.3+574451 (hereafter \textit{Swift} J1644+57) was initially detected by the \textit{Swift} Burst Alert Telescope (BAT) as a transient source. Subsequent observations by the \textit{Swift} X-ray Telescope (XRT) revealed a bright X-ray counterpart and confirmed the first unambiguous detection of a relativistic accretion-powered jet associated with a TDE. This event exhibited an intense X-ray flare for approximately three days, reaching a peak X-ray flux of $5\times10^{-9}~\rm erg~cm^{-2}~s^{-1}$ \citep{SwiftJ1644_Burrows_2011,SwiftJ1644_2058_Seifina_2017}.
    
    \item \textit{Swift} J2058.4+0516 (hereafter \textit{Swift} J2058+05) is the second X-ray TDE exhibiting similar relativistic jet characteristics to those observed in \textit{Swift} J1644+57 \citep{SwiftJ2058_Pasham_2015}.
    \end{enumerate}
    
\begin{table*}
    \caption{Ten TDEs with X-ray observations between 2008 and 2018.}
    \label{table:TDE}
    \resizebox{\textwidth}{!}{ 
    \begin{tabular}{lcccccccc}
    \hline
    \multirow{2}{*}{Discovery Name} & \multirow{2}{*}{MJD} & \multirow{2}{*}{$z$} & $\log_{10}L_{\mathrm{bol,~peak}}$ & $\log_{10}L_{\mathrm{X,~peak}}$ & $\log_{10}L_{\mathrm{IR,~peak}}$ & $\log_{10}F_{\mathrm{X,~peak}}$ & \multirow{2}{*}{$\log_{10}(M_{\mathrm{BH}}/M_\odot)$} & \multirow{2}{*}{TS} \\
     &  &  & (erg s$^{-1}$) & (erg s$^{-1}$) & (erg s$^{-1}$) & (erg cm$^{-2}$ s$^{-1}$) &  &  \\
    \hline
    ATLAS17jrp$^1$         & 58156 & 0.066 & 44.15 & 43.10 & 43.32 & $-11.91$ & $6.55\pm0.45$ & 9.99 \\
    ASASSN-14li$^2$        & 56997 & 0.020 & $\sim44$ & 43.99 & 41.48 & $-10.00$ & $6.77\pm0.46$ & 2.42 \\
    ASASSN-15oi$^3$        & 57324 & 0.048 & 44.11 & 42.69 & 41.63 & $-12.01$ & $6.42\pm0.48$ & 0.57 \\
    OGLE16aaa$^4$          & 57548 & 0.165 & 44.58 & 43.69 & --     & $-12.19$ & $7.14\pm0.48$ & 2.61 \\
    PS18kh$^5$             & 58220 & 0.071 & 43.94 & 41.74 & 41.82 & $-13.35$ & $5.83\pm0.51$ & 0.00 \\
    SDSS J1201+30$^6$      & 55357 & 0.146 & 44.48 & 44.48 & --     & $-11.05$ & $6.77\pm0.46$ & 0.02 \\
    2MASX 0740-85$^7$      & 56748 & 0.017 & 44.30 & 42.52 & --     & $-11.31$ & $6.77\pm0.46$ & 0.06 \\
    XMMSL2 J1446+68$^8$    & 57622 & 0.029 & $\sim43$ & 42.78 & --     & $-11.77$ & $8.38\pm0.53$ & 0.00 \\
    \textit{Swift} J1644+57$^9$ & 55648 & 0.354 & --    & 48.46 & --     & $-8.30$ & $5.50\pm1.10$ & 0.00 \\
    \textit{Swift} J2058+05$^{10}$ & 55698 & 1.185 & --  & 47.48 & --     & $-10.10$ & $\sim7.70$ & 0.00 \\
    \hline
    \end{tabular}
    } 
    \vspace{5pt}
    
    \raggedright 
    \textbf{Notes.} Columns: (1) Discovery name of the TDE; (2) Date of the first X-ray detection (MJD); (3) Redshift of the host galaxy; (4) Bolometric peak luminosity; (5) X-ray peak luminosity; (6) Infrared peak luminosity corresponds to the dust emission modeled as a blackbody; (7) X-ray peak flux, calculated from the reported luminosity and redshift, assuming a flat $\Lambda$CDM cosmology with $H_0=70~\mathrm {km ~s^{-1}~Mpc^{-1}}$, $\Omega_M=0.3$, and $\Omega_\Lambda=0.7$; (8) Black hole mass estimated via the $M_{\rm{BH}}\text{-}\sigma_\star$ relation based on the velocity dispersion of the host galaxy; (9) Test statistic value from the time-dependent neutrino search conducted in this work. Source data are compiled from \citet{X-ray_TDE_Saxton_2021}, \citet{IR_TDE_Jiang:2021klj}, \citet{Guolo:2023bds} and \citet{Wang&Jiang_2022_ATLAS17jrp}. 
\end{table*}

\section{Analysis and Results}\label{sec:analyse}
\subsection{The Time-dependent Analysis method}\label{subsec:method}
    We analyze the public muon-track events using the maximum likelihood method to search for time-dependent neutrino emission from the target source. The data are modeled as a two-component mixture of signal and background \citep{Braun_2008}. The likelihood function is the product of the probability densities for all muon events within a $5^\circ$ circle centered on the target source, defined as the region of interest.
    
    \begin{equation}
        \mathcal{L}(n_{\mathrm{s}})=\prod_{i}^{N}\left(\frac{n_{\mathrm{s}}}{N}S_i+(1-\frac{n_{\mathrm{s}}}{N})B_i\right),
        \label{eq:like}
    \end{equation}
    where $n_{\mathrm{s}}$ denotes the number of signal neutrinos and $N$ is the total number of muon events in the data sample. The signal and background probability density functions (PDFs) for the $i$-th event are given by,
    \begin{equation}
	S_i=S^{\text{spat}}_i(\vec{x}_i,\sigma_i|\vec{x}_s)   S^{\text{ener}}_i(E_i|\vec{x}_s,\gamma) S^{\text{temp}}_i(t_i|T_0,T_{\mathrm{W}}),
    \end{equation}
    \begin{equation}
	B_i=B^{\text{spat}}_i(\delta_i) B^{\text{ener}}_i(E_i|\delta_i) B^{\text{temp}}_i(t_i).
	\label{eq:bpdf} 
    \end{equation}
    Here, $S^{\text{spat}}_i$ is the spatial PDF of the signal, with $\vec{x}_i$ and $\vec{x}_s$ denoting the locations of the $i$-th event and the source, respectively, and $\sigma_i$ is the reconstructed angular uncertainty of the $i$-th event. The spatial PDF of the signal is modeled as a two-dimensional Gaussian spatial distribution \citep[as shown in Equation (7) in][]{LiRonglan_2022}. $S^{\text{ener}}_i$ is the signal energy PDF, describing the probability density distribution of energy $E_i$ with an assumed spectral energy distribution (SED) \citep[as shown in Equation (6) in][]{Huang_MN_2022}. We assume that the neutrino spectrum follows a power law distribution, $dN_\nu /dE_\nu\propto E_\nu^{-\gamma}$, where $E_\nu$ is the real energy of neutrinos, and $\gamma$ is the spectral index. We adopt a single power-law neutrino spectrum as a generic, model-independent search template widely used in IceCube analyses. Given the limited statistics of potential transient signals, introducing additional spectral parameters is not supported by the data and may lead to overfitting. This choice ensures a robust search strategy without imposing a specific physical emission model. The signal temporal PDF, $S^{\text{temp}}_i$, describes the distribution of events over time, is written as
    \begin{equation}
	S_i^{\text{temp}}(t_i|T_0,T_{\mathrm{W}})=\frac{1}{T_{\mathrm{W}}},\;\;\;T_0-T_{\mathrm{W}}/2<t_i<T_0+T_{\mathrm{W}}/2,
	\label{eq:stime}
    \end{equation}
    where we employ a box-shaped temporal PDF parameterized by central time $T_0$ and full width $T_{\mathrm{W}}$, with events uniformly distributed within this temporal window.
    Compared to a time-integrated search, the time-dependent search may reduce the background and enhance the probability of identifying neutrino sources. In Eq.~$\left(\ref{eq:bpdf}\right)$, $B^{\text{spat}}_i$ represents the background spatial PDF, and $B^{\text{ener}}_i$ is the background energy PDF. Both are nearly uniform in right ascension, with detailed descriptions provided in \cite{Huang_MN_2022}. The background temporal PDF is assumed to be constant, i.e., $B^{\text{temp}}=1/T_{\mathrm{L}}$, where $T_{\mathrm{L}}$ is the livetime of the data sample.

        \begin{figure*}
        \centering
        \includegraphics[width=\textwidth]{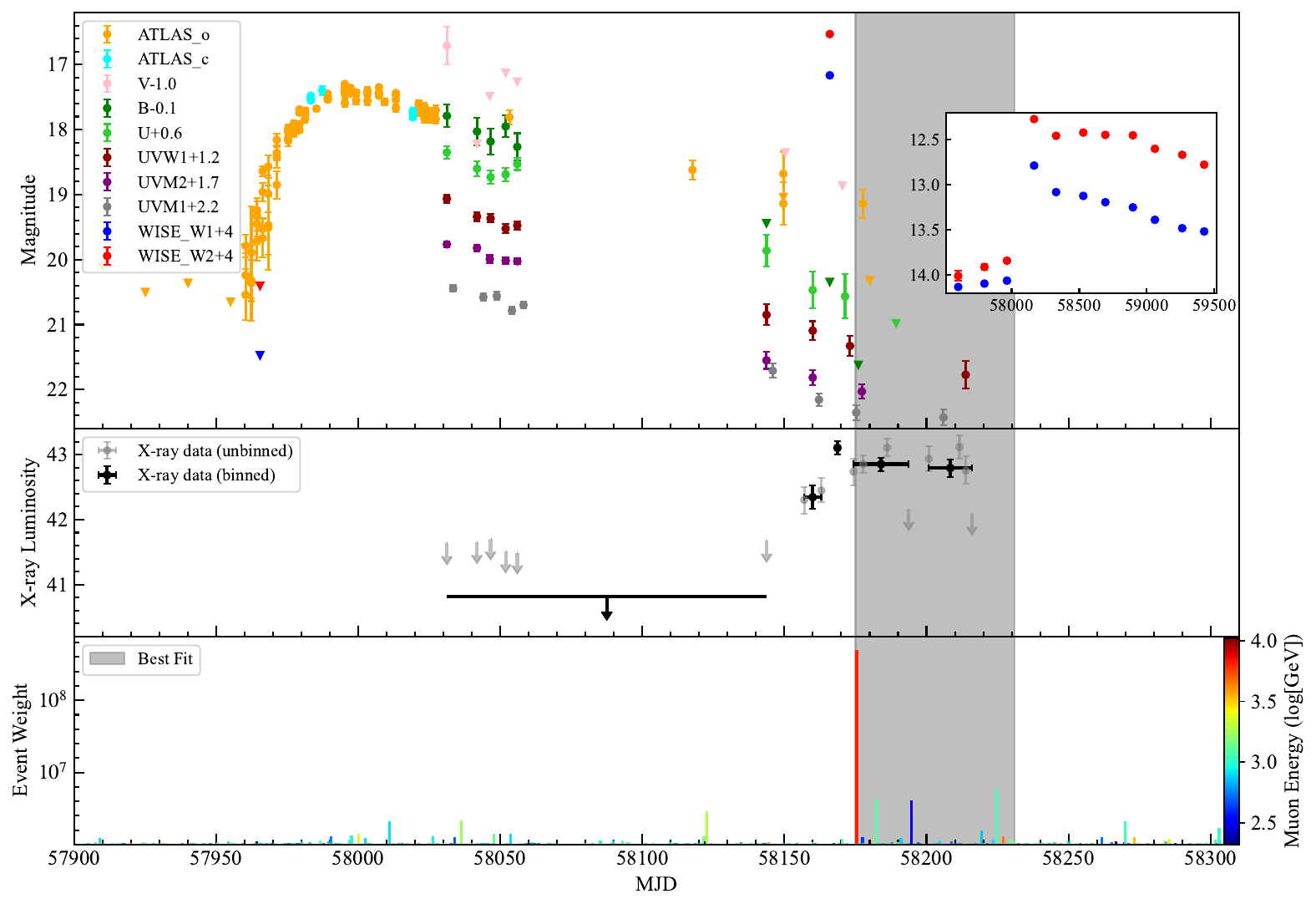}
        \caption{
            \textbf{Top panel}: Host-galaxy-subtracted light curve of ATLAS17jrp in the optical, UV, and mid-IR bands. The inset shows the long-term mid-IR light curve from WISE without host-galaxy subtraction. 
            \textbf{Middle panel}: The gray points represent the 0.3-2 keV X-ray luminosities observed by \textit{Swift}-XRT, while the black points represent the binned X-ray light curve, with the error bars indicating the binned regions. 
            \textbf{Bottom panel}: Time-independent weight of individual events during the multi-band radiation period of ATLAS17jrp. Each vertical line's length represents the weight value of an event observed at a specific MJD, corresponding to the $S_i^{\mathrm{spat}}S_i^{\mathrm{ener}}/B_i^{\mathrm{spat}}B_i^{\mathrm{ener}}$ of IceCube events. The color of each event indicates the reconstructed muon energy (log[GeV]), from which the neutrino energy of a single event can also be inferred. The gray shaded region indicates the best-fitting neutrino flare window.
        }
        \label{fig:all}
    \end{figure*}
	
\subsection{Searching for the neutrino emission}\label{subsec:analysis}
    Following the methodology described in Section~\ref{subsec:method}, we conducted a search for neutrino cluster signals during the electromagnetic radiation phases of these TDEs. We use a likelihood ratio test to compare the null and alternative hypotheses. The null hypothesis corresponds to the background-only model, denoted $\mathcal{L}\left(n_{\mathrm{s}}=0\right)$. The test statistic (TS) is defined as twice the log-likelihood ratio 
    \begin{equation}
	TS=2\log \left [\frac{T_{\mathrm{W}}}{T} \times \frac{\mathcal{L}(n_{\mathrm{s}},\gamma,T_0,T_{\mathrm{W}})}{\mathcal{L}(n_{\mathrm{s}}=0)} \right ],
	\label{eq:ts}
    \end{equation} 
    where the additional factor $T_{\mathrm{W}}/T$ corrects for the look-elsewhere effect \citep{Braun_2010,IceCube:txs0506} due to choosing a time window of width $T_{\mathrm{W}}$ from the duration $T$, defined as the electromagnetic radiation time of the TDE.
    The alternative hypothesis incorporates four parameters for each source at position $\vec{x}_s$: the number of signal events $n_{\mathrm{s}}\ge0$, the spectral index $1.0\le \gamma \le 4.0$, the central time $T_0$ constrained to the interval $\left[{T_{\mathrm{min}}}+T_{\mathrm{W}}/2,~{T_{\mathrm{max}}}-T_{\mathrm{W}}/2\right]$ (where ${T_{\mathrm{min}}}$ denotes the source discovery time and $T_{\mathrm{max}}$ is defined as the earlier of the source's observation end time and the IceCube data cutoff time), and the time window width $T_{\mathrm{W}}<T$ with $T=T_{\mathrm{max}}-T_{\mathrm{min}}$. We maximize the test statistic over these parameters to identify the best-fit neutrino burst.
	
    To investigate the temporal correlation between electromagnetic emission and high-energy neutrinos in TDEs, the time-dependent analysis was conducted during the electromagnetic emission phase of these TDEs. We present the TS values corresponding to the best-fit neutrino flare signals from these sources in Table~\ref{table:TDE}. The neutrino flare associated with ATLAS17jrp shows the highest TS value among these sources. The potential neutrino flare with the best-fitting parameters being $\hat{n}_\mathrm{s}=8.3$, $\hat{\gamma}=2.7$, $\hat{T}_\mathrm{0}=58203$ (MJD), and $\hat{T}_\mathrm{W}=56$ days. 
    The significance of the association is determined by the events' spatial proximity to ATLAS17jrp, their temporal clustering, and their energies. 
    In the bottom panel of Fig.~$\ref{fig:all}$, we plot the time-independent weights of individual events in the likelihood analysis during the multiwavelength emission phase of ATLAS17jrp. Each weight corresponds to the ratio of the product of spatial and energy terms in the signal PDF to the corresponding product in the background PDF, calculated for the best-fit spectral index of $\gamma=2.7$. The corresponding OUV, IR, and X-ray light curves are displayed in the top and middle panels for comparison. The best-fit neutrino flare occurred 19 days after the onset of X-ray emission and 6 days after the X-ray peak time. Subsequent analysis will focus on probing the correlation between neutrino emission and X-ray activity in ATLAS17jrp. 
	
    Based on the fitting results, we can calculate the corresponding $\nu_\mu+\bar{\nu}_\mu$ neutrino flux within the neutrino flare window. This calculation is based on the function of $n_{\mathrm{s}}$ as described in Eq.~$\left(\ref{eq:ns}\right)$, which represents the number of signal neutrinos predicted by the model.
    \begin{equation}
	n_{\mathrm{s}} = T_{\mathrm{W}}\times \int A_{\mathrm{eff}}(E_\nu,\delta_i)\frac{dN_\nu}{dE_\nu}(E_\nu)\;dE_\nu,
	\label{eq:ns}
    \end{equation}
    where $T_{\mathrm{W}}$ denotes the duration of neutrino flare, and $A_{\mathrm{eff}}$ represents the effective area of the detector at declination $\delta_i$. Using the $1\sigma$ error range obtained from the fit $\hat{n}_{\mathrm{s}}=8.3\pm3.8$ and $\hat{\gamma}=2.7\pm0.4$, we derived the flux normalization of muon neutrino spectrum at 100 TeV, with $\Phi_{\mathrm{\nu_\mu+\bar{\nu}_\mu}}^{\mathrm{100\;TeV}}=6.2_{-5.6}^{+24}\times 10^{-19}~\mathrm{GeV^{-1}~cm^{-2}~s^{-1}}$. We multiply the muon neutrino flux by a factor of 3 to estimate the all-flavor neutrino flux, assuming equal fluxes for all flavors after oscillation. In Fig.~\ref{fig:spec_nu}, we show the best-fitting results and the uncertainty intervals using the green solid line and shaded area. 
    
    In addition, we performed a time-integrated search using the 10-year of IceCube muon-track data centered on the position of ATLAS17jrp. The analysis yielded a null result, with a best-fit signal number of $n_s \approx 0$ and a TS of 0, which indicates no significant steady-state emission. This result confirms the unique value of time-dependent analysis. Unlike time-integrated searches that average out brief signals, the time-dependent approach can identify short bursts that would otherwise be hidden in the background. Thus, for transient sources like TDEs, a time-dependent search is essential for identifying neutrino bursts. 
    
\subsection{Probability by Chance}\label{subsec:p_value}
\begin{figure}
        \centering
        \includegraphics[width=0.45\textwidth]{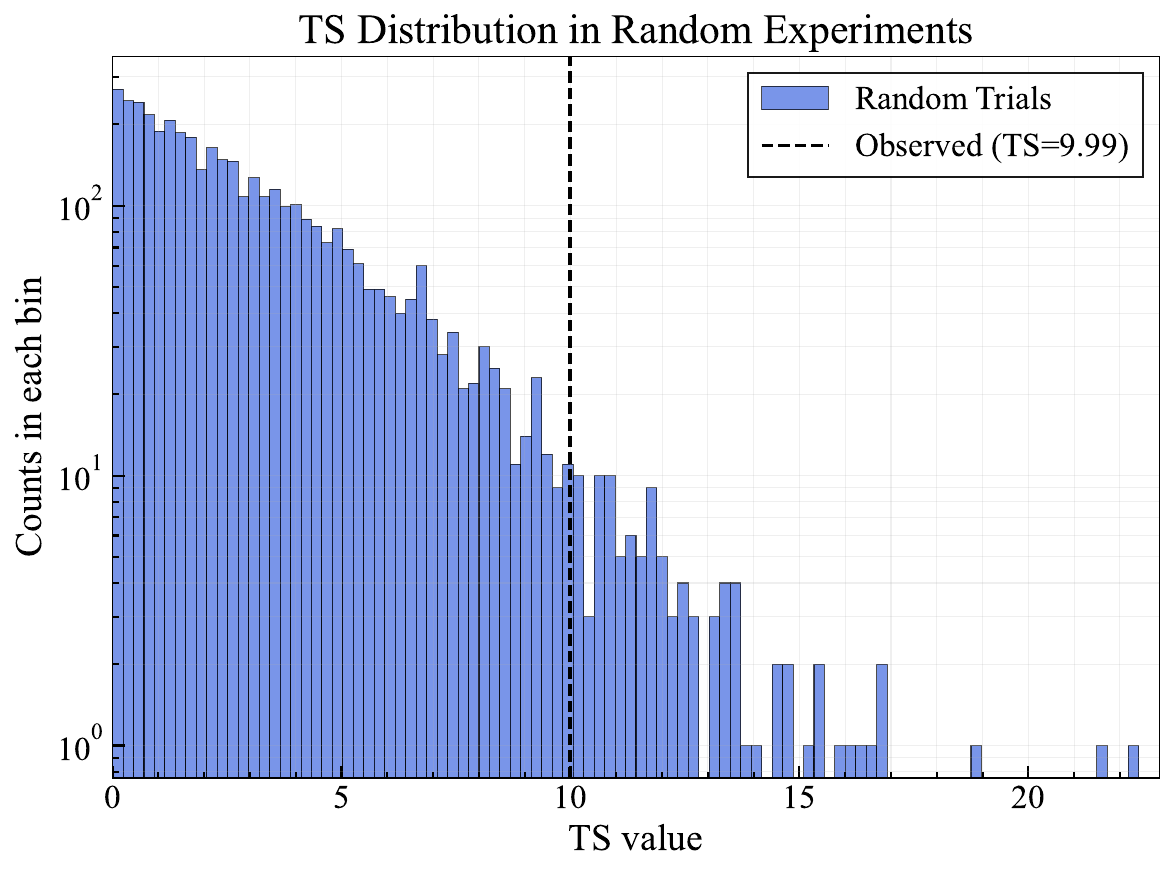}
        \caption{Time-dependent analysis test statistic distributions. The purple histogram shows the distribution of $TS$ values derived from 10,000 scrambled trials in the time-dependent analysis. The black dashed line marks the $TS$ value observed at the location of ATLAS17jrp in the time-dependent analysis.}
        \label{fig:ts_dist}
    \end{figure}

    To assess the significance of the neutrino flare associated with ATLAS17jrp, we performed 10,000 null hypothesis simulations by shuffling the right ascension and event times within the IC86-II-VII sample, which covers the electromagnetic observation period of ATLAS17jrp. For each simulation, the maximum TS at the coordinates of ATLAS17jrp was calculated following the analysis described in Section~\ref{subsec:analysis}. The resulting distribution of the maximum TS from these 10,000 null hypothesis simulations is presented in Fig.~\ref{fig:ts_dist}. This figure illustrates the probability of observing a given TS value from chance spatial fluctuations alone. From the simulations, we found that 107 trials (out of 10,000) had a maximum TS value larger than or equal to the observed TS in the real data, i.e., $TS\geq 9.99$. This corresponds to a pre$\text{-}$trial $p\mathrm{\text{-}value}$ of 0.011 for the spatial association. However, to establish a genuine temporal correlation, a more stringent condition is required. When we further required a temporal coincidence (a delay of less than 19 days) between the neutrino flare and the X-ray emission, only 11 simulations with $TS\geq 9.99$ satisfied the temporal criterion, yielding a pre$\text{-}$trial $p\mathrm{\text{-}value}$ of 0.0011. To account for the look-elsewhere effect from searching over 10 X-ray TDEs, we applied a trial correction using $p_\mathrm{post}=1-\left(1-p_\mathrm{pre} \right)^N$ with $N=10$. The final post-trial $p\text{-value}$ is 0.01, representing the chance probability of detecting such a neutrino flare temporally correlated with the X-ray emission.

\section{Multi-messenger modeling of ATLAS17jrp}\label{model}
    \begin{figure*}
    \centering
    \includegraphics[width=0.85\textwidth]{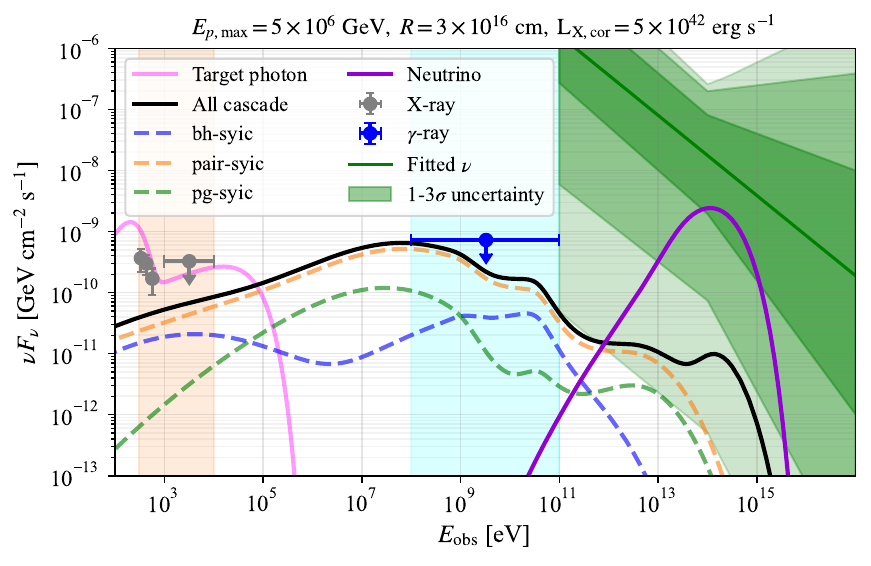}
        \caption{Multi-messenger SEDs of ATLAS17jrp. The pink, black, and purple curves represent the spectra of the input target photon, the electromagnetic cascade, and all-flavor high-energy neutrino emission produced by our model, respectively. The `bh-syic', `pair-syic', and `pg-syic' components represent the synchrotron and inverse Compton emissions from electrons and positrons produced by the Bethe-Heitler (BH), $\gamma\gamma$, and $p\gamma$ processes, respectively. The grey points represent the X-ray data at the peak time. The blue line indicates the $\gamma$-ray upper limit set by \textit{Fermi}-LAT. The green solid line shows the best-fit neutrino spectrum obtained through data fitting, with shaded regions representing the $1\sigma$, $2\sigma$, and $3\sigma$ uncertainty regions derived from the same analysis.} %
	\label{fig:spec_nu}
    \end{figure*}
    
    Given the temporal proximity of neutrino and X-ray flares, we consider the X-rays as target photons interacting with accelerated protons to investigate multimessenger emissions from the spherically symmetric radiation region. In Fig.~{\ref{fig:spec_nu}}, the orange shaded area indicates the observation energy range of 0.3-10 keV for the \textit{Swift}-XRT instrument. \textit{Swift}-XRT detected only soft X-ray photons from ATLAS17jrp. The gray data points represent the X-ray flux at the peak time (\textit{Swift}-XRT observation ID: 00010335009\footnote{\url{https://www.swift.ac.uk/archive/browsedata.php?oid=00010335009&source=obs&reproc=1}}), modeled assuming a power-law spectrum in the 0.3-1 keV range. The upper limit above 1 keV is defined as the flux that exceeds the background by $5\sigma$. The soft X-ray emission observed with \textit{Swift}-XRT can be well described by a thermal blackbody spectrum, consistent with emission originating from the inner region of the accretion disk. The temperature and luminosity of the soft X-ray photons are derived by fitting the observational data with a blackbody model. We describe the temporal behavior of the injected target photons following the X-ray light curve shown in Fig.~\ref{fig:all}. In addition to the observed soft X-ray emission, we assume that the corona can generate additional hard X-ray emission by following the approach used for AGNs \citep{Murase:2019vdl}. We describe the hard component of the X-ray spectrum produced by the corona as a power law with an exponential cutoff, as $E^2dN_{X, \rm cor}/dE\propto E^{-\Gamma_{\rm cor}}\exp(-E/E_{X,\rm cut})$. In this expression, the spectral index and the cutoff energy can be determined by $\Gamma_{\mathrm{cor}}\approx 0.167\log(\lambda_{\mathrm{Edd}})$ \citep{Trakhtenbrot:2017xiz} and $E_{X,\rm cut}\approx[-74\log(\lambda_{\mathrm{Edd}})+150]$ keV \citep{Ricci:2018eir}, where $\lambda_{\rm Edd}\approx 10L_{X,\rm cor}/L_{\rm Edd}$ and $L_{\rm X,\rm cor}=\int dE (EdN_{X,\rm cor}/{dE})$ is the corona luminosity. In this approach, the coronal X-ray spectrum can be determined solely by the luminosity $L_{X,\rm cor}$. The coronal X-ray luminosity is estimated to be $L_{X,\rm cor}\lesssim5\times10^{42}~\rm erg~s^{-1}$ under the constraint of the observed upper limit provided by \textit{Swift}-XRT in the 1-10 keV range.
    
    Protons can be accelerated in the compact inner jet, the accretion disk or corona, or an extended isotropic wind \citep{Murase:2020lnu} with a magnetic field strength of $B=$ 0.1-1 G, similar to the case of active galactic nucleus (AGNs). For the sake of generality, we do not specify the exact site of particle acceleration. We assume a power-law injection rate for the accelerated protons, i.e. $Q_p\propto E_p^{-2}\exp\left( -E_p/E_{p,\rm max}\right)$, where $E_{p,\rm max}$ is the maximum proton energy. We normalize the injection spectrum using $L_p$ and the radiation zone volume $V=4\pi R^3/3$, with $R$ representing the radius of the radiation zone, i.e. $\int E_pQ_pdE_p=L_p/V$. For proton injection, we follow the approach of \cite{Winter_2023} and \cite{Yuan_2023} by treating the injected proton luminosity as a fraction of the accretion power, $L_p=\epsilon_p\dot M_{\rm{BH}}c^2$, where $\epsilon_p=0.2$ is a fiducial value. During the peak accretion phase, the TDE can exceed the Eddington limit, with the ratio of the peak accretion rate $\dot M_{\rm BH}(t_{\rm opt,peak})$ to the Eddington accretion rate $\dot M_{\rm Edd}$ being $\gtrsim$1-10 \citep{Dai:2018jbr}, where $t_{\rm opt,peak}$ is the peak time of the OUV emission. Explicitly, we define 
    \begin{equation}
	\zeta=\frac{\dot M_{\rm BH}(t_{\rm opt,peak})}{\dot M_{\rm Edd}}=\frac{\dot M_{\rm BH}(t_{\rm opt,peak})}{L_{\rm Edd}/(\eta_{\rm rad}c^2)}\gtrsim 1\text{-}10,
    \end{equation}
    where $L_{\rm Edd}\approx 1.3\times10^{44} (M_{\rm BH}/10^7M_\odot)~\rm erg~s^{-1}$ is the Eddington luminosity and the radiation efficiency $\eta_{\rm rad}\sim0.01\text{-}0.1$ \citep{McKinney:2015lma} represents the fraction of accreted power reprocessed in the radiation. We normalize the proton injection luminosity at $t_{\rm opt,peak}$ via 
    \begin{equation}
        \begin{split}
            L_p(t_{\rm opt,peak})&=\epsilon_p(\zeta/\eta_{\rm rad})L_{\rm Edd} \\
            &=1.2\times10^{46}~\rm erg~s^{-1}~\left(\frac{\epsilon_p}{0.2}\right) \left(\frac{\zeta/\eta_{\rm rad}}{100}\right) \left(\frac{M_{\rm BH}}{10^7~M_\odot}\right),
	\end{split}
    \end{equation}
    where $\zeta/\eta_{\rm rad}=100$, as used by \cite{Winter_2023}. We assume that the proton injection luminosity and accretion rate follow the $t^{-5/3}$ decay of the OUV light curve, which reflects the mass fallback onto the SMBH. 
    
    After protons are injected into the radiation zone, they interact with target thermal photons ($p\gamma$ process) or target protons ($pp$ process), losing energy and producing secondary particles, which include charged/neutral pions, muons, neutrinos, and electrons. Explicitly, we estimate the free-escaping optical thickness for $p\gamma$ and $pp$ interactions
    \begin{equation}
	\begin{split}
		\tau_{p\gamma}&\approx\frac{\sigma_{p\gamma}L_X}{4\pi Rc\varepsilon_X}\\
		&\sim8.3\times10^{-2}\left(\frac{R}{10^{16}~\rm cm}\right)^{-1}\left(\frac{\varepsilon_X}{100~\rm eV}\right)^{-1},
	\end{split}
    \end{equation}
    and 
    \begin{equation}
        \begin{split}
		\tau_{pp}&\approx\frac{\sigma_{pp}\eta_w\dot M_{\rm BH}}{4\pi Rm_pv_w}\\
            &\lesssim9.3\times10^{-4}\left(\frac{v_w}{0.1c}\right)^{-1}\left(\frac{R}{10^{16}~\rm cm}\right)^{-1}\left(\frac{M_{\rm BH}}{10^7~M_\odot}\right),
	\end{split}
    \end{equation}
    where $\sigma_{p\gamma}\simeq 500~\mu \rm b$ is the $p\gamma$ cross section, $L_X$ is the bolometric X-ray luminosity, $\varepsilon_X$ is the peak energy of thermal X-rays, $\sigma_{pp}\simeq40{~\rm mb}$ is the $pp$ cross section at 1 PeV, $\eta_w\lesssim 0.1$ is the fraction of accreted mass that is converted to winds, and $v_w\sim0.1~c$ is the typical disk wind velocity. Noting that $\tau_{p\gamma}/\tau_{pp}\gtrsim100$, we conclude that the $p\gamma$ interactions dominate the production of neutrino and electromagnetic (EM) cascade emissions. Assuming the magnetic field strength of $B\sim0.1$ G, neutral pions, charged leptons, and the electron/positron pairs generated from $\gamma\gamma$ annihilation and Bethe-Heitler (BH) process will deposit the energy into radiation, initiating the electromagnetic cascades. 
	
    In this work, we assume that the proton acceleration radius is smaller than the dust radius, consistent with the scenario proposed in \cite{Winter_2023}. The dust radius for ATLAS17jrp is estimated in \cite{MN_AT2017gge}, and based on these considerations, we set the radius of the radiation zone to be $3\times10^{16}~\rm{cm}$. Incorporating the time-dependent target photon spectra (pink curve in Fig.~\ref{fig:spec_nu}) and the proton injection rate, we adopted the following model parameters as $R=3\times10^{16}~\rm{cm}$, $B=0.1$ G, $L_{X,\rm cor}=5\times10^{42}~\rm erg~s^{-1}$, and $E_{p,\rm max}=5\times 10^6$ GeV. We use the open-source Astrophysical Multi-Messenger Modeling \citep[AM$^3$,][]{AM3_2023} software to solve the time-dependent transport equations for all relevant particle species\footnote{A detailed description of the particle energy loss and diffusion terms can be found in \cite{Yuan_2023}}. We evolve the system to the end time of the neutrino flare, resulting in the average energy spectra of the neutrino and electromagnetic cascade over the period, shown as the purple and black solid curves in Fig.~\ref{fig:spec_nu}. No GeV counterpart was detected by \textit{Fermi}-LAT \citep{Fermi_4FGLDR4}, leading to an upper limit of $\mathrm{1.35\times 10^{-12}\;erg~cm^{-2}~s^{-1}}$ in the range of 0.1-100 GeV, derived from the 14-year detection threshold map\footnote{\url{https://fermi.gsfc.nasa.gov/ssc/data/access/lat/14yr_catalog/detthresh_P8R3_14years_PL22.fits}}. 

    Our data analysis, based on a single power-law spectral hypothesis, yields a best-fit neutrino signal of $n_s = 8.3$ with a soft spectral index ($\gamma = 2.7$). In contrast, the spectrum predicted by our one-zone isotropic model is harder and gives an expected count of $n_{s,\rm model} \simeq 0.085$. The total expected event count is only consistent with the data within the $3\sigma$ uncertainty range (0.002-45.68), but Fig.~\ref{fig:spec_nu} shows that around 100 TeV the flux predicted by this model is consistent with the observation within $1\sigma$ range. The discrepancy between the model and the data likely originates from lower energies, indicating that our model may be missing an additional source of low-energy neutrinos. A compact coronal region is a natural candidate for this missing component, analogous to the neutrino-emitting corona of Seyfert Galaxy NGC 1068, in which a dense X-ray field would enhance $p\gamma$ interactions and yield a softer neutrino spectrum. Simulations show that a turbulent TDE corona can produce delayed neutrino emission (peaking near 10 TeV) while its inherent $\gamma\gamma$ absorption satisfies GeV gamma-ray limits \citep{Yuan:2025ctq}. The broader connection between coronal neutrinos and X-ray emission is further supported by recent multi-messenger models \citep{Murase:2020lnu, Yuan:2025ctq, wangxj2025}, which independently argue for coronae as viable neutrino production sites in transient systems.

\section{Discussion and Conclusion}\label{conclusion}
    In this study, we performed a time-dependent analysis of 10-year IceCube muon-track data and observed a potential neutrino flare spatially and temporally coincident with the TDE ATLAS17jrp. The neutrino flare occurred 19 days after the onset of the X-ray emission and lasted for approximately 56 days. The chance probability of detecting a neutrino flare at ATLAS17jrp during its electromagnetic emission period, with a significance exceeding that of the observed signal, is 0.01 (pre\text{-}trial, for the spatial association). When further requiring that the flare onset occurs within 19 days of the X-ray emission, this probability is reduced to 0.001 (pre\text{-}trial, for the spatial and temporal association), reflecting the temporal correlation between the neutrino and X-ray emission. To account for the search for over ten X-ray-detected TDEs, we applied a trial correction for multiple tests. The resulting post\text{-}trial $p$\text{-}value of 0.01 accounts for the look-elsewhere effect inherent in the source catalog search. Given that this significance is modest, the post-trial result does not constitute a firm detection. Rather, it indicates a suggestive correlation that warrants further investigation.
    
    To interpret the potential signal, we consider neutrino production via proton-photon ($p\gamma$) interactions between relativistic protons and X-ray photons, which serve as the target field. Proton-proton ($pp$) interactions are not included, as they are expected to be subdominant in isotropic TDE winds \citep{Yuan_2023}. Interactions with IR photons are also neglected, since they predominantly generate neutrinos above 10 PeV, beyond the observed energy range. We model the production of neutrinos and electromagnetic cascades in TDEs via interactions between relativistic protons and X-ray photons, where the X-ray emission serves as the target photon field. Based on the assumption of specific parameters, we present the corresponding electromagnetic cascade and neutrino spectra in Fig.~\ref{fig:spec_nu}. Our model can produce a neutrino flux at energies of several hundred TeV without violating the $\gamma$-ray upper limit observed by \textit{Fermi}-LAT. However, accounting for the entire potential signal, particularly at lower energies, might require contributions from other regions, such as the hot corona or the inner accretion disk. Although our study is constrained by the limited sample size, it highlights the potential of TDEs as neutrino sources and underscores the need for continued high-cadence X-ray monitoring and next-generation neutrino observatories to test this connection.
    
    Furthermore, more precise measurements of neutrino energy spectra, combined with correlations inferred from multi-wavelength observations, will enhance our understanding of the underlying radiation mechanism. These efforts are crucial for uncovering the physical processes driving high-energy astrophysical processes and for identifying the origins of astrophysical neutrinos. Neutrino telescopes both currently in operation and under construction, including the Cubic Kilometer Neutrino Telescope \citep[KM3NeT;][]{KM3Net:2016zxf}, IceCube-Gen2 \citep{IceCube-Gen2:2020qha,IceCube-Gen2:2021fkg}, the Giant Radio Array for Neutrino Detection \citep[GRAND;][]{GRAND_scienc}, the Tropical Deep-sea Neutrino Telescope \citep[TRIDENT;][]{TRIDENT_NA}, the High-energy Underwater Neutrino Telescope \citep[HUNT;][]{HUNT:2023mzt}, and the NEutrino Observatory in the Nanhai \citep[NEON;][]{NEON}, are expected to improve sensitivity and increase the detection rate of high-energy neutrinos. The X-ray survey by Einstein Probe \citep[EP;][]{EP_2025} and the transient monitoring capabilities of the Space-based multi-band astronomical Variable Objects Monitor \cite[SVOM;][]{SVOM} will improve the identification of TDEs with X-ray emission, thereby expanding the statistical sample available for analysis. When combined with multi-messenger observations, these efforts will contribute to identifying the origins of astrophysical neutrinos and advancing time-domain multi-messenger astronomy.

\section*{Acknowledgements}
    H.N.He is supported by NSFC under grants No. 12321003, No. 12173091, No. 12333006, Project for Young Scientists in Basic Research of Chinese Academy of Sciences (No. YSBR- 061), and the Strategic Priority Research Program of the Chinese Academy of Sciences No.XDB0550400. 
    Y.W is supported by the Jiangsu Funding Program for Excellent Postdoctoral Talent (grant No. 2024ZB110), the Postdoctoral Fellowship Program (grant No. GZC20241916) and the General Fund (grant No. 2024M763531) of the China Postdoctoral Science Foundation. 

\section*{Data Availability}
    The 10-year (2008–2018) muon-track data by IceCube are available at \url{https://icecube.wisc.edu/data-releases/2021/01/all-sky-point-source-icecube-data-years-2008-2018/}. 
    The Swift-XRT observation of ATLAS17jrp (ObsID: 00010335009) can be accessed at \url{https://www.swift.ac.uk/archive/browsedata.php?oid=00010335009&source=obs&reproc=1}. 
    The multiwavelength data points of ATLAS17jrp shown in Fig.~\ref{fig:all} are adopted from \citet{Wang&Jiang_2022_ATLAS17jrp}, available at \url{https://iopscience.iop.org/article/10.3847/2041-8213/ac6670}. Information on X-ray TDEs is available through the references listed in Section~\ref{subsec:TDE} and Table~\ref{table:TDE}.

\bibliographystyle{mnras}
\bibliography{ATLAS17jrp}

\bsp	
\label{lastpage}
\end{document}